\documentclass[aps,prd,reprint,nofootinbib,
showpacs,showkeys,superscriptaddress]{revtex4-1}

\usepackage{graphicx}
\usepackage{epstopdf}
\usepackage{amsmath}
\usepackage{amssymb}
\usepackage{amsfonts}
\usepackage{latexsym}
\usepackage{amssymb}
\usepackage{amsfonts}
\usepackage{latexsym}
\usepackage{subfigure}
\usepackage{float}
\usepackage{color}
\usepackage{amsthm}
\usepackage{graphicx}
\usepackage{color}
\usepackage{hyperref}

\newcommand{\nn}{\nonumber\\}

\renewcommand{\vec}[1]{{\bf #1}}

\def\beq{\begin{eqnarray}}
\def\eeq{\end{eqnarray}}
\def\nn{\nonumber }                                 
\def\r{\ref}                                                

\def\diag{\,\mbox{diag}\,}

\def\be{\beta}

\def\de{\delta}

\def\ep{\epsilon}

\def\la{\lambda}

\def\rho{\rho}
\def\si{\sigma}
\def\om{\omega}

\def\Om{\Omega}

\def\QG{ quantum gravity }
\def\B1{ Bianchi-I }

\def\GW{ gravitational wave }
\def\HL{ Hubble - Lema\^{\i}tre }

\renewcommand{\vec}[1]{{\bf #1}}

\renewcommand{\diag}{\,\mbox{diag}\,}

\begin{document}

\title{Beyond the linear analysis of stability in higher derivative gravity
\\
with the Bianchi-I metric}
\vskip 2mm

\author{Simpliciano Castardelli dos Reis}
\email{simplim15@hotmail.com}
\affiliation{Departamento de F\'{\i}sica, ICE,
Universidade Federal de Juiz de Fora
\\
Campus Universit\'{a}rio - Juiz de Fora, 36036-330, MG, Brazil}

\author{Grigori Chapiro}
\email{grigorichapiro@gmail.com}
\affiliation{Departamento de Matem\'{a}tica, ICE,
Universidade Federal de Juiz de Fora
\\
Campus Universit\'{a}rio - Juiz de Fora, 36036-330, MG, Brazil}

\author{Ilya L. Shapiro}
 \email{shapiro@fisica.ufjf.br}
\affiliation{
 Departamento de F\'{\i}sica, ICE, Universidade Federal de Juiz de Fora
\\
Campus Universit\'{a}rio - Juiz de Fora, 36036-330, MG, Brazil}
\affiliation{
Tomsk State Pedagogical University, Tomsk, 634041, Russia}
\affiliation{National Research Tomsk State University,
Tomsk, 634050, Russia.}

\date{\today}

\begin{abstract}
\noindent
The study of stability of gravitational perturbations in higher derivative
gravity has shown that at the linear level the massive unphysical ghost is
not generated from vacuum if the initial seed of metric perturbation has
frequency essentially below the Planck  threshold. The mathematical
knowledge indicated that the linear stability is supposed to hold even
at the nonperturbative level, but in such a complicated case it is
important to perform a verification of this statement. We compare the
asymptotic stability solutions at the linear and full nonperturbative
levels for the Bianchi-I metric with small anisotropies, which can be
regarded as an extreme, zero frequency limit of a gravitational wave.
As one should expect from the combination of previous analysis and
general mathematical theorems, there is a good correspondence between
linear stability and the nonperturbative asymptotic behavior.
\end{abstract}



\maketitle

\section{Introduction}
\label{S1}

There is well known controversy between renormalizability of \QG and
the problems which are caused by the introduction of higher derivatives,
which are capable to provide this renormalizability \cite{Stelle}. The
theory with sufficiently general higher derivatives always has massive
unphysical ghosts in the spectrum, making physical interpretation of
such a theory problematic. In the presence of ghosts the vacuum state
is not stable, and even Minkowski space may decay into Planck-mass
ghost plus the gravitons with huge overall energy which is
compensating the negative energy of the ghost.

Indeed, the presence of the ghost in the spectrum of the theory does
not necessary mean that there should be such a particle ``alive''. It
might happen, e.g., that there is an unknown physical principle
which forbids the concentration of gravitons with Planck energy
density, resolving the mentioned puzzle with Minkowski space
\cite{HD-Stab,DG}, and
also providing the stability of a qualitatively similar, low curvature
space-times. Certain arguments which support this expectation
have been given in the recent papers \cite{GW-Stab,HD-Stab,
GW-HD-MPLA}. In a perfect agreement with the previous works
on the evolution of gravitational waves on the deSitter background,
\cite{star81,asta,hhr}, we have found that these waves do
not have growing amplitudes, regardless of the presence of higher
derivatives. The situation was analysed in the context of ghosts in
\cite{HD-Stab}, where it was shown that there are no growing modes
also in other cosmological backgrounds, if the initial frequency of
the gravitational wave is much smaller than the Planck scale. On the
opposite, in case of  Planck-scale frequencies there is an expected
explosion of gravitational waves. Our interpretation of this situation
in \cite{HD-Stab} was that the presence of the ghost in the spectrum
of the theory does not necessary means that there is a ghost as a
real particle. For the low-energy frequencies of the gravitational
waves the positive energy modes don't form a Planck-density
distribution and then the ghost can not be created from vacuum. This
solution of the problem is certainly incomplete, because {\it i)}
\QG is supposed to work at all frequencies, even over-planckian
ones;  \ \ {\it ii)} The linear stability guarantees non-linear
perturbative stability from the mathematical point of view, but
it does not look sufficient from the point of view of Physics,
because the exponential instabilities are expected at the
non-linear level \cite{Woodard}.

The item {\it i)} has been addressed in \cite{PP},  where
we have shown that, at least for the cosmological background, if the
cosmological solution corresponds to the rapidly expanding universe,
the explosive behaviour of the gravitational waves does not last for
long, and after that the metric perturbations get stabilized. The reason
is that the wave equation includes the wave vector ${\vec k}$ only in
the combination ${\vec q}=\textstyle{\frac{\vec k}{a(t)}}$, such that
the physical frequency of the wave is decreasing as $1/a$. Of course
this is not the complete solution of the problem, but just a useful
hint on how the problem can be eventually solved. What is still
needed is certainly the physical principle explaining why gravitons
can not accumulate with the over-Planck energy density on a weak
gravitational background, and how this principle may be violated by
the fast expansion of the universe.

In the present work we address the point {\it ii)} and check out
whether the situation with stability changes when we go beyond
the linear perturbations level. In fact, we are able to get even the
non-perturbative results, but not for the usual gravitational waves.
Instead, we shall consider the evolution of anisotropies in the
framework of the Bianchi I cosmological metric.
Since the pioneering work \cite{Kasner:1921zz}, the Bianchi I
metric have been extensively studied as a model of anisotropic
homogeneous cosmology. For cosmologic solutions and stability
in fourth derivative gravity, see recent works \cite{Barrow:2006xb,Toporensky:2016kss,Muller:2017nxg}.

With respect to an arbitrary perturbations of the metric our
approach means the following two restrictions: (a) small
amplitude of the perturbations; (b) zero frequencies of the
perturbations. In what follows we perform numerical analysis
of the dynamics of anisotropies under these two assumptions.

The paper is organized as follows. In the next Sec. \ref{S2} the
equations for the \B1 metric in the fourth derivative gravity
are derived in Misner parametrization
\cite{Misner:1967uu,Misner:1969hg}. Before starting the numerical
analysis of the full and linearized version of these equations, in
Sec.~\ref{Smath} we present a brief survey of the mathematical
knowledge on the subject of stability in the systems described by
differential equations. Namely, we discuss to which extent the
stability with respect to linear perturbations defines the behavior
of the system at the nonperturbative. In Sec.~\ref{S3} we present
the results of numerical analysis including comparison of linear
and full versions of equations. Finally, in Sec.~\ref{S4} we draw
our conclusions and discuss possible extensions of the present work.

\section{Dynamical equations}
\label{S2}

The theory of our interest has the classical action
\beq
S = \int d^4x\,\Big(-\frac{M_P^2}{16\pi}\,R
+ a_1 C^2 + a_2 R^2\,\Big).
\label{action}
\eeq
Here $M_P$ is the Planck mass, while other parameters $a_{1}$ and
$a_{2}$ are arbitrary dimensionless constants. $R$ and $C^2$ are,
respectively, the Ricci scalar and the square of Weyl tensor,
\beq
C^2= R_{\mu\nu\alpha\beta}^2 - 2 R_{\alpha\beta}^2 + \frac{1}{3}R^2.
\nn
\eeq
According to the recent work \cite{Barrow:2006xb}, every vacuum
solution 
of Einstein field equations is also a solution of the theory
(\ref{action}). However, since there are higher derivatives, the
theory (\ref{action}) can develop strong instabilities which are not
present in general relativity. These instabilities represent our main
interest in what follows.

In a comoving and synchronous frame, the \B1 anisotropic metric is
\beq
ds^2 = dt^2 - a^2_{1}(t)\,dx^2-a_{2}^2(t)\,dy^2-a_{3}^2(t)\,dz^2.
\label{lineelement}
\eeq

One can switch to a more useful parametrization, introduced by Misner
in \cite{Misner:1967uu,Misner:1969hg}, in which there is a separation
between the functions of time responsible for \textit{expansion}
$\si(t)$ and \textit{shear} of the universe $\be_{\pm}(t)$ respectively,
\beq
a_{1}(t) &=& e^{\si}\, e^{\be_{+}+ \sqrt{3} \be_{-}},
\nonumber
\\
a_{2}(t) &=& e^{\si}\, e^{\be_{+}- \sqrt{3} \be_{-}},
\nonumber
\\
a_{3}(t) &=& e^{\si}\, e^{-2\beta_{+}}.
\label{misnerparametrization}
\eeq
In what follows the term \textit{anisotropies} will refer to the
functions $\beta_{\pm}$. The trivial case $\beta_{\pm}=0$
corresponds to an isotropic metric. A usefulness of Misner
parametrization resides in the possibility of perform a local
conformal transformation
\beq
g_{\mu\nu} = e^{2\si(\eta)}\,\bar{g}_{\mu\nu},
\label{conformaltransformation}
\eeq
where the conformal time $\eta$ is defined by the relation
$\,dt=e^{\si(\eta)}d\eta$. The fiducial metric $\bar{g}_{\mu\nu}$
is given by (\ref{misnerparametrization}) with $\,\si(t)\equiv 0$.
Under a conformal transformation, the Weyl-squared part of
the action (\ref{action}) is expressed only in terms of the metric
$\bar{g}_{\mu\nu}$, while Ricci scalar transforms as
\beq
R = e^{-2\si}\big[\bar{R} - 6 (\si')^2 - 6\si''\big].
\label{riccitransform}
\eeq
It is easy to check that $\sqrt{-\bar{g}}=1$ and the expressions
for $\bar{R}$ and $\bar{C}^2$ are
\beq
\bar{R} &=& -6\,(\,\beta'_{+}{}^2 + \beta'_{-}{}^2\,),
\nonumber
\\
\bar{C}^2
&=&
12 \big(\beta''_{+}{}^2 + \beta''_{-}{}^2\big)
\,+\, 48\big(\beta'_{+}{}^2 + \beta'_{-}{}^2\big)^2
\nonumber
\\
&+& 16\big[\beta'_{+}\big(3\beta'_{-}{}^2
- \beta'_{+}{}^2 \big)\big]'.
\label{ricciandweyl}
\eeq
In these expressions the prime stands for the derivative with
respect to conformal time.

Let us remember that we regard the anisotropy parameters as a
truncated part of the gravitational wave, or the \GW with zero
frequency. The \GW of our interest is supposed to be created
by quantum fluctuations \cite{GW-Stab}, and if it does not
experience fast growth due to the presence of ghosts, its amplitude
remains very small. This is our {\it main assumption} and we need to
know whether it is violated by the dynamics of the gravitational
wave or, in the truncated case, of the anisotropies. Thus, consider
the physically most interesting case when the anisotropy parameters
in Eq.~(\r{misnerparametrization}) are small,
$\,\vert \be_\pm \vert \ll 1$.
Then one can write the space components of the metric in the form
\beq
&&
g_{ik}\,=\,-\,\de_{ik}\,+\,h_{ik},
\nn
\\
&&
h_{ik}\,=\,-\,\diag
\big(\be_{+} + \sqrt{3}\be_{-} ,\,\,
        \be_{+} - \sqrt{3}\be_{-},\,\, -2\be_{+} \big).
\mbox{\qquad}
\label{weak}
\eeq
It is easy to see that the trace of the last expression is zero,
$\,\de^{ik} h_{ik}=0$, exactly as in the case of the gravitational
wave, also in both cases we have two degrees of freedom.

Another desired similarity would be a transverse nature of
the wave. However, in the case of \B1 metric this feature can not
be verified, because the perturbation in (\ref{weak}) is dependent
only on time, and there is no wave vector. Therefore there is no
complete correspondence between   (\ref{weak})  and the
gravitational wave, and we can speak only about a qualitative
similarity between the two types of the perturbations. At the same
time, since the Ostrogradsky instabilities which are expected in the
higher derivative theories \cite{Ostrog} (see \cite{Woodard:2015zca}
for a recent review) appear due to the higher derivatives in time,
we can expect that the data obtained by using \B1 metric will provide
a useful hint for the general situation with the stability of metric
perturbations in the higher derivative theories. Since the wave
vector is zero in the case of  (\ref{weak}), we can expect that,
according to the results of \cite{HD-Stab}, the classical isotropic
solutions will be stable in the linear approximation. The \B1
metric offers a possibility to have an independent check of these
results and, most relevant, to go beyond the linear approximation.

In terms of the new variables, discarding superficial terms and
taking into account that in \B1 case all metric components depend
only on time and not on the spatial coordinates, the Lagrangian of
the action (\ref{action}) becomes
\beq
\mathcal{L}
&=&
- \,\frac{3\,M_P{}^2}{8\,\pi}\,e^{2\si}\,\big[\si'{}^2
- \big(\be'_{+}{}^2 + \be'_{-}{}^2\big)\big]
\\
&+&
12 \big(3 a_2+4 a_1)\,(\be'_{+}{}^2 + \be'_{-}{}^2\big)^2
+ 12a_1\big(\be''_{+}{}^2 + \be''_{-}{}^2\big)
\nn
\\
&+&
72 a_2\big(\si''+ \si'{}^2)\,(\be'_{+}{}^2 + \be'_{-}{}^2\big)
+ 36 a_2\big(\si''+ \si'{}^2\big)^2.
\nonumber
\label{lagrangian}
\eeq
It is worth noting that in the limit of general relativity $a_{1,2} \to 0$
and after a rescaling anisotropies, that doesn't affect the dynamics of the
conformal factor, we recover the conventional Lagrangian for the
gravitational waves beyond the horizon
\cite{Mukhanov:2005sc,Gorbunov:2011zzc}. This means that,
at least in the linear order, the \B1 model under consideration
can be seen as a zero-frequency approximation of the equation
for the gravitational waves. Thus we shall assume that this
correspondence holds beyond the linear order and regard
the \B1 as a simplest version of the equation for the
gravitational wave.

It is easy to see that that Lagrangian expression has terms which
are second and fourth order in conformal time derivatives. It is
useful to show explicitly the unit of time $\eta_0$.
The dynamical equations can be obtained by taking the variational
derivatives of the action with the Lagrangian (\ref{lagrangian}). The
presence of isotropically distributed matter, radiation or
cosmological constant does not affect the equations for
$\,\be _{\pm}$ \cite{Reis:2017bjf,hervik2007}, but only changes the
equation for $\sigma$ through the trace of the energy-momentum
tensor. We will only consider a perfect fluid with linear equation of
state defined by the constant $\omega$ which is assuming the values
$\frac{1}{3}$, $0$ and $-1$ for radiation, dust and cosmological
constant, respectively. Taking variational derivatives with respect
to $\sigma(\eta)$ and $\be_{\pm}(\eta)$ and adding the matter
part, we arrive at the equations
\beq
&&
72a_2\,\Big[\si^{(4)} - 2\,\si''\,\big(3\,\si'^2
+ \,\beta _-'{}^2  + \beta_+'{}^2\big)
\nonumber
\\
&-&
4 \si'\,\big(\beta _-' \beta _-''  +\beta_+' \beta_+''\big)
\nonumber
\\
&+&
2\,\big(\,\beta _-' \,\beta _-^{\prime\prime \prime}
+ \beta _+'\,\beta _+^{\prime \prime \prime}\,\big)
+ 2 \big(\beta _-''{}^2 +  \be_+''{}^2\big)\Big]
\nonumber
\\
&+&
\frac{3}{4 \pi }e^{2\si}\,M_p^2\,\eta_{0}^2
\Big[\big(\be_-'{}^2+\be _+'{}^2+\si''+\si'^2\big)
\nonumber
\\
&-&
\frac12\,(1-3\omega) e^{(1-3\om)\si}
\Big]\,=\,0
\label{sigmaequation}
\mbox{\quad}\mbox{\quad}
\eeq
and
\beq
&&
24 a_1
\big(8\be_{\mp}'{}^2\, \beta _{\pm}''
+ 16\beta _{\pm}'\,\beta _{\mp}'\,\beta_{\mp}''
+ 24\beta_{\mp}'{}^2\, \beta_{\pm}''
- \beta_{\pm}{}^{(4)}\big)
\nonumber
\\
&+&
\frac{3}{4\pi}\,e^{2\sigma}\,M_p^2 \eta_{0}^2
\,\Big(\be_{\pm}''+2\si' \,\be_{\pm}' \Big)
\nonumber
\\
&&
+144 a_2
\Big[\beta_{\pm}'
\big(2\si'\,\si''+ 2\, \beta_{\mp}'\,\beta_{\mp}''
+\si^{\prime\prime\prime}\,\big)
\nonumber
\\
&+&
\beta_{\pm}''\,\big(\,\si'^2+ 3\beta_{\pm}'{}^2
+ \beta_{\mp}'{}^2+ \si''\big)
\Big] \,=\,0.
\mbox{\quad}\mbox{\quad}
\mbox{\quad}\mbox{\quad}
\label{betasequation}
\eeq
Here the primes mean the derivative with respect to the conformal
time measured in the units of $\,\eta_0$.
Eq.~(\ref{sigmaequation}) corresponds to the variation with
respect to $\,\sigma$ with the perfect fluid contribution,
where $\Om_{0}$ is the relative energy density of matter or
cosmological constant. The sum of  $\Om$ and the contribution
of higher derivative terms is equal to one identically. The
Eqs.~(\ref{betasequation})
describe the nonlinear dynamics of anisotropies.

We can also express the dynamical equations in terms of physical
time through the relation $\,dt = e^{\si(\eta)}d\eta$. The results are
\beq
&&
72a_{2}\,\Big[\,\si^{(4)}
+ 12\dot{\si}^2\,\ddot{\si}+4\,\ddot{\si}^2
\nonumber
\\
&+&
\dot{\si}\,(\,6\,\dot{\be}_+ , \ddot{\be}_+
+ 6\dot{\be}_-\,\ddot{\be}_- + 7\,\si^{(3)}\,)
\nonumber
\\
&+&
2 \,\big(\,\ddot{\be}_+{}^2 +\ddot{\be}_-{}^2
+ \dot{\be}_+ \, \be_{+}^{(3)}
+ \dot{\be}_- \,\be_{-}^{(3)}\big)\,\Big]
\nonumber
\\
&+&
\frac{3}{4\pi}\,\Big(\frac{M_p}{H_{0}}\,\Big)^2
\,\Big[\,2\,\dot{\si}^2+\dot{\be}_+^2 + \dot{\be}_-^2
\nonumber
\\
&-&
2\Omega_{\Lambda}
-\frac{1}{2}\,\Omega_{0}\,e^{-3\si(1+\om)}\,(1-3\om)\Big]\,=\,0
\label{sigmaequationphy}
\eeq
and
\beq
&&
144 a_2
\Big\{
\ddot{\be}_{\pm}\,\big(\,2\,\dot{\si}^2
+\ddot{\be}_{\mp}^2
+3\,\dot{\be}_{\pm}^2 + \ddot{\si}\,\big)
\nonumber
\\
&+&
\dot{\be}_{\pm}
\Big[
6\,\dot{\si}^3
+ 3\dot{\si}\,(\,\dot{\be}_+^2 + \dot{\be}_+^2\,)
+ 7\,\dot{\si}\,\ddot{\si}+2\,\dot{\be}_{\mp}\ddot{\be}_{\mp}
+\si^{(3)}
\Big]\Big\}
\nonumber
\\
&+&
24a_{1}\,\Big\{\dot{\be}_{\pm}
\Big[6\,\dot{\si}^3
-16\,\dot{\be}_{\mp}\,\ddot{\be}_{\mp} + \si^{(3)}
\nonumber
\\
&+&
7\,\dot{\si}\,\ddot{\si}-24\,\dot{\si}\big(\,\dot{\be}_+^2
+ \dot{\be}_-^2 \big)\Big] + 6\dot{\si}\be_{\pm}{}^{(3)}
\,+ \,\be_{\pm}{}^{(4)}
\nonumber
\\
&+&
\ddot{\be}_{\pm}\,\big(11\,\dot{\si}^2-8\,\dot{\be}_{\mp}^2
- 24 \dot{\be}_{\pm}^2+4\,\ddot{\si}\big)\Big\}
\nonumber
\\
&+&
 \frac{3}{4\pi}
\Big(\frac{M_p}{H_{0}}\,\Big)^2\,\Big(\ddot{\be}_{\pm}
+ 3\dot{\si}\dot{\be_{\pm}}\Big)\,=\,0.
\label{beatasequationsphy}
\eeq
Here the dots mean derivative with respect of dimensionless time
$\tau = H_{0}\,t$, where $H_{0}$ is the Hubble - Lema\^{\i}tre
parameter measured at some instant of time.
The set of Eqs.~(\ref{sigmaequationphy}) and (\ref{betasequation}) or
(\ref{sigmaequationphy}) and (\ref{beatasequationsphy}) represent
systems of three coupled ordinary differential equations of the fourth
order.

Besides the Einstein space solutions in vacuum (with cosmological
constant), there are no much chances to find an exact solution of this
system, and this is not our purpose in the complicated case with
higher derivative terms included. Instead, we shall explore the
stability of the cosmological (homogeneous and isotropic) solutions,
corresponding to $\,\beta _{\pm}=0\,$ and the $\,\si(t)=\si_0(t)\,$
given by some cosmological solutions.

An important point concerns the choice of the background solution
$\,\si_0(t)$. Let us start from a few preliminary observations. The
{\it first} one is that in the action (\ref{action}) the term $a_1C^2$
(regardless being most relevant for the tensor perturbations and
massive ghosts) does not affect the dynamics of the the conformal
factor and therefore the one of $\,\si_0(t)$. Thus when we choose
this functions, we do not need to take the Weyl-squared term into
account. {\it Second}, our main target in the previous works on the
cosmological stability in the presence of massive ghosts was the
{\it low-energy} cosmological solutions. Thus, the canonical
approach would be to ignore also the $a_2R^2$-term as being
Planck-suppressed (the last means we consider such solutions for
which
$\,\left\vert a_2R^2 \right\vert \ll \left\vert M_P^2R \right\vert$
in the action and the corresponding hierarchy in the equations of
motion), and consider the classical radiation-dominated
and dust-radiation solutions only. Let us stress that this hierarchy
can be assumed {\it only} for the background $\,\si_0(t)$. For the
perturbations such as gravitational waves, the run-away solutions
are capable to easily destroy this hierarchy. The main subject of
the present work is to explore the effect of non-linearities in this
possible breaking in the framework of the simple Bianchi-I based
model.

The {\it third} point is that we can easily extend the low-energy
region for the background up to the inflation scale, just taking the
$a_2R^2$-term into account. According to the available set of
observational and experimental data, this term is the main
ingredient of the Starobinsky model \cite{star}, that is exactly the
most successful phenomenologically model of inflation. In order to
achieve this success \cite{star83}, the value of $a_2$ should be
chosen at
about $\,5 \times 10^8$. Then the inflationary solution corresponds
to the slowly decreasing Hubble parameter, with an approximately
linear dependence $H(t)$. Then, since our ultimate interest is the
dynamics of the gravitational waves with the initial frequencies
much greater in magnitude than $H$ (and at the same time much
below the Planck scale \cite{HD-Stab}), it is a very good
approximation to regard $H$ as a constant. Thus, we can safely
consider, instead of the linear $H(t)$, the constant $H$ and derive
it from the classical cosmological constant. All in all, we arrive at
the situation where the main features of our model can be
explored taking the three simplest examples of $\,\si_0(t)$, namely
cosmological constant-, radiation- and matter-dominated classical
solutions.

Let us repeat that the main advantage of the \B1 metrics is that
the Eqs.~(\ref{sigmaequation}) and  (\ref{sigmaequation}) or
(\ref{sigmaequationphy}) and  (\ref{beatasequationsphy}) are
relatively simple and can be explored numerically even at the
non-perturbative level. Thus we get a chance to check by direct
calculation whether the mathematical statements about the general
relation between linear stability and the nonperturbative asymptotic
behavior, which were used  in  \cite{HD-Stab} and \cite{PP}, are
correct. However,  before going to numerics we shall give a brief
survey of the mentioned mathematical statements in the next section.


\section{Asymptotic series expansion for singular perturbation}
\label{Smath}

Since our intention is to compare the linear approximation for the
anisotropies with the nonperturbative numerical solution, it makes
sense to briefly review the general mathematical theorems which
cover the relation between first order stability and nonperturbative
behavior in the systems described by differential equations.

In the zero-order case functions $\si$ and $\be_\pm$ are approximated
by $\si_0(t)$ and zero, because in the background solutions there are
no anisotropies, by assumption. This fact motivates to explore the
general solution of the system of equations
Eqs.~(\ref{sigmaequationphy}) and  (\ref{beatasequationsphy}) in
the form of asymptotic series expansion
\beq
&&
\dot\sigma = \sigma^0 + \epsilon \sigma^1 + \cdots
\nn
\\
&&
\dot\beta_\pm = 0 + \epsilon \beta^1_\pm + \cdots \,,
\label{eq:expansion}
\eeq
where $\ep$ is a small parameter, which one can easily implemented
into the perturbations (\ref{weak}).

Eqs.~(\ref{sigmaequationphy}) and  (\ref{beatasequationsphy})
can be rewritten in the mathematically standard form as a system of
twelve autonomous ordinary differential equations
\beq
d_t \mathbf{y} \, = \,
\frac{d}{dt}\mathbf{y} \, = \, \mathbf{f}(\mathbf{y}),
\label{veceq}
\eeq
where the vector $\mathbf{y}$ includes $\sigma$, $\beta_\pm$
and also first, second and third derivatives of these functions.
Substituting into this system the expansion \eqref{eq:expansion},
we arrive at the equations for the power series
\beq
d_t \big[
\mathbf{y}^0 + \epsilon \mathbf{y}^1 + \cdots\big]
\,=\, \mathbf{f}(\mathbf{y}^0)
+ \epsilon\nabla \mathbf{f}(\mathbf{y}^0) \mathbf{y}^1 + \cdots,
\mbox{\qquad}
\label{com}
\eeq
where $\nabla \mathbf{f}(\mathbf{y}^0)$ is a Jacobian of the function
$\mathbf{f}$ calculated on the background (unperturbed) solution
$\mathbf{y}^0$. In order to solve this system we equate terms with
the same order in $\ep$. This procedure is well known in Singular
Perturbation Theory \cite{wasow87}.

Let us note that the order zero in $\epsilon$ corresponds to the
equation $d_t \mathbf{y}^0 = \mathbf{f}(\mathbf{y}^0)$, that
is satisfied for the background under consideration. Then the first
order approximation corresponds to the linear differential equation
\beq
d_t \mathbf{y}^1 = \nabla \mathbf{f}(\mathbf{y}^0) \mathbf{y}^1.
\label{lin}
\eeq
Our main purpose is to compare the solution of this equation with
the one for the complete version (\ref{com}). For instance, let us
assume that for the certain choice of initial conditions (small
deviations from the background, as we explained above), linear
system (\ref{lin}) does not show growing modes, but only those
which asymptotically vanish or oscillate without growing amplitude
in the limit $t \to \infty$. Then,
under smoothness hypotheses on the dependence on the small
parameter $\ep$, the first order approximation
$\mathbf{y}^0 + \epsilon \mathbf{y}^1$ is of the order $\ep$
close to the solution of the complete system
$d_t \mathbf{y}  = \mathbf{f}(\mathbf{y})$ \cite{wasow87}.

Finally, we can quote the following two theorems  concerning sink
equilibrium points, which can be found in the well-known book on
differential equations \cite{hirsch1974differential}:
\vskip 2mm

\noindent
\textbf{Theorem 1.} \ \
Assume that the system $d_t \mathbf{y}  = \mathbf{f}(\mathbf{y})$
possesses a sink in the point $\tilde{\mathbf{y}}$, i.e., there exists a
constant $c>0$, such that all eigenvalues $\lambda_i$
of the Jacobian $\mathbf{f}(\tilde{\mathbf{y}})$
satisfy $Re(\la_i)<-c$. Then all the solutions starting
in some neighborhood of the point $\tilde{\mathbf{y}}$ converge
to $\tilde{\mathbf{y}}$ exponentially.
\vskip 2mm

\noindent
\textbf{Theorem 2.} \ \
If the system $d_t \mathbf{y}  = \mathbf{f}(\mathbf{y})$
possesses a stable equilibrium in $\tilde{\mathbf{y}}$, then all
eigenvalues $\lambda_i$ of the Jacobean
$\mathbf{f}(\tilde{\mathbf{y}})$ have non positive real part of
the eigenvalues $Re(\la_i)\leq 0$.
\vskip 2mm

Coming back to our problem of exploring Eqs.~(\ref{sigmaequationphy})
and  (\ref{beatasequationsphy}), we know that in the linear approximation
there are no growing modes for the frequencies below the Planck-order
threshold \cite{HD-Stab,GW-HD-MPLA}. This is certainly true for the
zero frequency modes, which correspond to the \B1 model. Thus we can
claim that the condition of the Theorem 2 are satisfied and, therefore,
the conditions of the Theorem 1 are also satisfied. Hence we can
expect a good qualitative correspondence between the dynamics of
anisotropies in the linear approximation and within the full
nonperturbative consideration. In the next section we check this
conclusion by using numerical methods and now let us discuss how
these well-known theorems can be applied to evaluate the regions
where one can expect the validity of the linear approximation.

First of all, let us construct the presentation (\ref{veceq}) for the
non-linear system formed by (\ref{sigmaequationphy}) and
(\ref{beatasequationsphy}). For this end we introduce the new
variables
\beq
&&
\dot{\si}\,=\,H,
\quad
\dot{H}\,=\,Q_1,
\quad
\dot{Q_1}\,=\,Q_2,
\nn
\\
&&
\dot{\be_{\pm}}\,=\,x_{\pm},
\quad
\dot{x_{\pm}}\,=\,y_{\pm},
\quad
\dot{y_{\pm}}\,=\,z_{\pm}
\label{newnonlin}
\eeq
Then the first order equations equations include (\ref{newnonlin}),
\beq
&&
\dot{Q_{2}}\,=\,-\Big[\,
12 H^2\,Q_1+4\,Q_{1}^2
\nn
\\
&&
+ H(6\,x_{+}\,y_{+}
+ 6 x_{-}\,y_{-}+7\,Q_{2}\,)
\nn
\\
&&
+ 2 \big(\,y_{+}^2 +y_{-}^2
+ x_{+}\,z_{+}
+ x_{-}\,z_{-}\big)\,\Big]
\nn
\\
&&
-\frac{M_p^2}{96\pi a_2 H_{0}^2}
\Big[2H^2+Q_{1} + x_{+}^2+x_{-}^2
- 2\Omega_{\Lambda}
\nn
\\
&&
-\frac{1}{2}\,\Omega_{0}\,e^{-3\si(1+\om)}\,(1-3\om)\Big],
\eeq
and
\beq
&&
\dot{z_{\pm}}\,=\,6\frac{a_2}{a_1}\,
\Big\{
x_{\pm}\,\Big[
6\,H^3
+ 3\,H\,(\,x_{+}^2+x_{-}^2\,)
+ 7\,H\,Q_{1}
\nn
\\
&&
+ 2x_{\mp}\,y_{\mp}
+Q_{2}\,\Big]+
y_{\pm}\,\big(\,2\,H^2
+x_{\mp}^2
+3\,x_{\pm}^2+Q_{1}\,\big)
\Big\}
\nonumber
\\
&&
- \Big\{
x_{\pm}
\big[6\,H^3
-16x_{\mp}\,y_{\mp}+Q_{2}
+ 7H\,Q_{1}
-24 H (x_{+}^2
+ x_{-}^2)\big]
\nonumber
\\
&&
+6\,H\,z_{\pm}
+ y_{\pm}\,\big(11\,H^2-8\,x_{\mp}^2
- 24 x_{\pm}^2+4\,Q_{1}\big)\Big\}
\nonumber
\\
&&
+\,\frac{3M_p^2}{4\pi a_1 H_{0}^2}
\big(y_{\pm}
+3H\,x_{\pm}\big).
\label{nonlinearsystem}
\eeq

The  first order version of linearized system consists from
\beq
&&
\dot{\be_{\pm}}\,=\,x_{\pm},
\quad \dot{x_{\pm}}\,=\,y_{\pm},
\quad \dot{y_{\pm}}\,=\,z_{\pm},
\eeq
and
\beq
&&
\dot{z_{\pm}}=\frac{6a_2}{a_1}
\Big\{x_{\pm}\big[
6\dot{\si_0}^3
+ 7\dot{\si_0}\ddot{\si_0}
+\si_0^{(3)}\big]
+ y_{\pm}\big(2\dot{\si_0}^2 + \ddot{\si_0}\big)
\Big\}
\nonumber
\\
&&
- \Big\{
x_{\pm} \big[6\dot{\si_0}^3
+\si_0^{(3)} + 7\dot{\si_0}\ddot{\si_0}\big]+6\,\dot{\si_0}\,z_{\pm}
\nonumber
\\
&&
+ y_{\pm}\,\big(11\,\dot{\si_0}^2+4\,\ddot{\si_0}\big)\Big\}
\,+\,\frac{3}{4\pi a_1\,H_{0}^2}
\,M_p^2\,\Big(\,y_{\pm}
+3\,\dot{\si_0}\,x_{\pm}\Big).
\label{linearsystem}
\eeq

In order to estimate the radius of the region where the linearization
procedure is valid for the ordinary differential equations written in
the form (\ref{veceq}), one needs to go into details of the proofs
of the theorems mentioned above. In both cases the proofs are
based on the Taylor expansions around the equilibrium point
$\mathbf{y_0}$ in the form
\beq
\mathbf{y}'
&=& (\mathbf{y_0}+\delta \mathbf{y})'
\,=\,
\mathbf{f}(\mathbf{y_0})
+ \mathbf{J} 
\delta \mathbf{y}
\nn
\\
&+&
\frac12(\delta \mathbf{y})^T \mathbf{H}
\delta \mathbf{y} + O((\delta \mathbf{y})^3),
\eeq
where $\mathbf{J}$ and $\mathbf{H}$ are the Jacobian and Hessian
operators of the function $\mathbf{f}$ evaluates on the background
 solution $\mathbf{y_0}$. Remember that at the equilibrium point $\mathbf{f}(\mathbf{y_0})=\mathbf{0}$ by definition. Thus, the
 equation above can be rewritten for the perturbations as
\beq
(\delta \mathbf{y})'
= \mathbf{J}(\mathbf{y_0}) \delta \mathbf{y}
+ \frac12 (\delta \mathbf{y})^T \mathbf{H}(y_0) \delta \mathbf{y}
+ O((\delta \mathbf{y})^3).
\mbox{\quad}
\label{Hess}
\eeq
The theorems cited above are valid in the region where the terms
of the higher order are negligible (or possibly vanish under certain
change of variables) in a small neighborhood of $\mathbf{y_0}$.
This means that the linear approximation ceases validity when linear
and quadratic terms are of the same order of magnitude. A rough
estimate for the region where the linear approximation is valid is
\beq
|\delta \mathbf{y}| < R, \qquad \text{where} \qquad
R = \mathcal{O}\Big(
\frac{||\mathbf{J}(\mathbf{y_0})||}{||\mathbf{H}(\mathbf{y_0})||}
\Big),
\label{chapiro_radius}
\eeq
where the Euclidean norm $|\cdot|$ is used for vectors and the
operator norm $\|\cdot\|$ follows the standard definition and can
be calculated using Riesz representation (see e.g. \cite{kreyszig78})
as
\beq
&&
\mathbf{J}|| = \max_{|\mathbf{y}|=1} |\mathbf{J}(\mathbf{y_0}) y|,
\label{J}
\nn
\\
\mbox{and}
\qquad
&&
\mathbf{H}|| =
\max_{|\mathbf{y}|=1;
|\mathbf{z}|=1} |\mathbf{y}^T \mathbf{H}(\mathbf{y_0}) \mathbf{z}|.
\label{eq:normas}
\eeq
One can note that it is not possible to apply this formula to the
linearized model as the Hessian tensor will be singular. This is a
natural situation, because the criteria (\ref{J}) and (\ref{eq:normas})
are intended to compare linear and non-linear cases.

In order to calculate the radius given in \ref{chapiro_radius} for
the system (\ref{nonlinearsystem}) we need just to evaluate the norms
of Jacobian and Hessian operators at the equilibrium point. The
numerical simulations based on Eqs.~(\ref{chapiro_radius}) with
(\ref{J}) and (\ref{eq:normas}) has been performed in the radiation
and dust models, using the dimensionless units with $M_P=1$. The
results were equal for the tested versions with $a_1=\pm 1$ and
$a_2=5\times 10^8$. In both models we met the radius
$R=(1/3)\cdot 10^{-9}$. It is interesting that the sign of $a_1$ did
not make any difference for the radius of validity of the linear
approximation $R$, regardless of the critical importance of the
same sign for the asymptotic stability, as we will discuss in the
next section.

An interesting observation is in order.  After we submitted the
first version of this work to arXiv we learned
about a similar investigation \cite{Salvio-19}. The results of
numerical analysis in this work concern the non-linear case and
are qualitatively the same as ours, that are also close to those of
the earlier paper \cite{Toporensky:2016kss}, which did not link
the study of the dynamics of anisotropies with the problem of
massive ghosts in
higher derivative gravity. The correspondence between the three
independent investigations are certainly adding an extra safety to
our conclusions. At the same time the results of \cite{Salvio-19}
include the growing solution for the  the initial conditions with
relatively large first derivatives of the anisotropies. This output
may look as a contradiction with our interpretation of \B1
perturbations as a zero frequency gravitational wave. The
analysis presented in this section shows that this is not a
correct interpretation. The frequency is still zero, but in this
case we have the situation when the initial conditions correspond
to the point which is out of the region satisfying the condition
(\ref{chapiro_radius}). As we have discussed, out of this region
we can not expect correspondence between linear and non-linear
approximations.

\section{Linear and non-linear numerical solutions}
\label{S3}

In this section we present the numerical solutions of differential
equations (\ref{sigmaequationphy}) and (\ref{beatasequationsphy})
in both linear and full version. The first part requires the linearization.
Let us note that in this section we exclusively work with set of
Eqs.~(\ref{sigmaequationphy}) and (\ref{beatasequationsphy})
in terms of dimensionless physical time.

\begin{figure}[H]
\centering
 \subfigure[fig1a][$\sigma$ solutions]
{\includegraphics[width=0.5\textwidth]{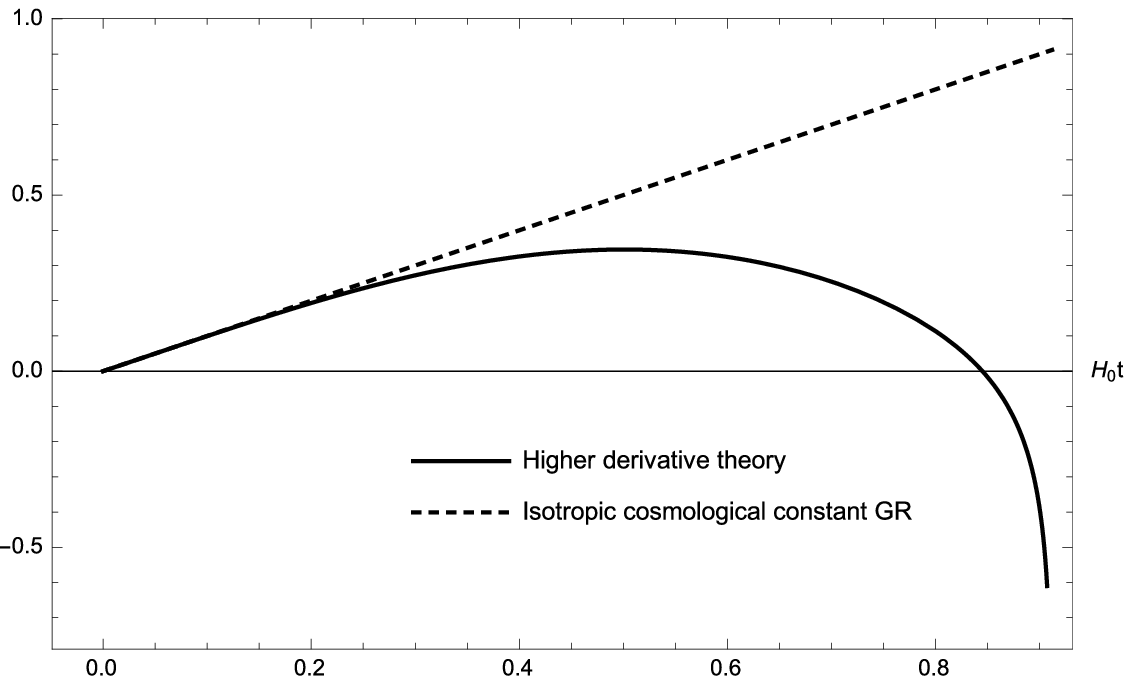}}
\subfigure[fig1b][anisotropies]
{\includegraphics[width=0.5\textwidth]{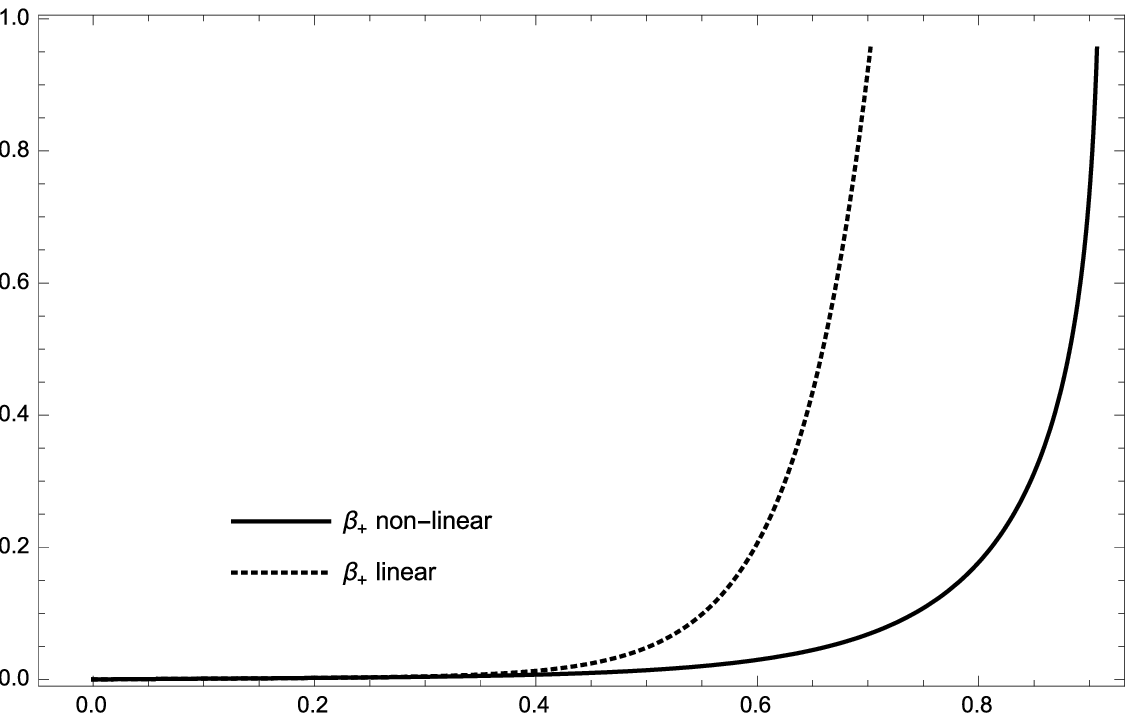}}
\begin{quotation}
\caption{Plots for $\,a_{1}=+1\,$ and $\,a_{2}=1\,$ in the cosmological
constant dominated case. For the anisotropies one can observe the
instability which is typical for the tachyonic ghost case for the more
general gravitational wave case \cite{HD-Stab,GWprT}.}
\label{a1positive-CC}
\end{quotation}
\end{figure}

As we have explained above, the linearization is performed around
isotropic cosmological solutions, which means null values for
anisotropies and the well-known cosmological solutions of general
relativity $\,\si_0(\tau)\,$. It is easy to check that at the linear
level the perturbations for $\,\si(\tau)\,$ and anisotropies
completely decouple. Thus in the linearized case one can restrict
consideration by the equations for anisotropies, which have the form
\beq
&&
\ddot{\be_{\pm}}
\bigg[(11 a_1-12a_2)\dot{\si_0}^2 + 2(2a_1-3a_2)\ddot{\si_0}
- \frac{3}{4\pi}\,\Big(\frac{M_p}{H_{0}}\Big)^2 \bigg]
\nonumber
\\
&&
+ 3\dot{\be_{\pm}}
\bigg[
8 \Big(a_1-6\,a_2)(6\dot{\si_0}^3
+7\dot{\si_0}\ddot{\si_0}+\si_{0}^{(3)}\Big)
\nn
\\
&&
- \frac{3}{4\pi}\Big(\frac{M_p}{H_{0}}\Big)^2\dot{\si_0}\bigg]
\,+\,
24a_1 \Big[\be_{\pm}^{(4)}+ 6\dot{\si_0}\be_{\pm}^{(3)}\Big]
\,=\,0.
\eeq
The free parameters of the systems are
\HL parameter at the reference time instant $H_0$ and the
coefficients $\,a_1\,$ and $\,a_2$. The theory with $\,a_1\,>\,0\,$
manifest instabilities for anisotropies as we know from the more
general gravitational wave solutions \cite{HD-Stab} (see also more
detailed discussion in \cite{GWprT}). For the same of completeness
we present the corresponding plots in Figs.~\ref{a1positive-CC} for
the cosmological constant - dominated and matter - dominated
backgrounds. 
The radiation case is very similar to these two and
hence will not be included here.
In what follows, we consider only negative values of $\,a_1$.
\begin{figure}[H]
\centering
 \subfigure[fig1a][$\sigma$ solutions]
{\includegraphics[width=0.5\textwidth]{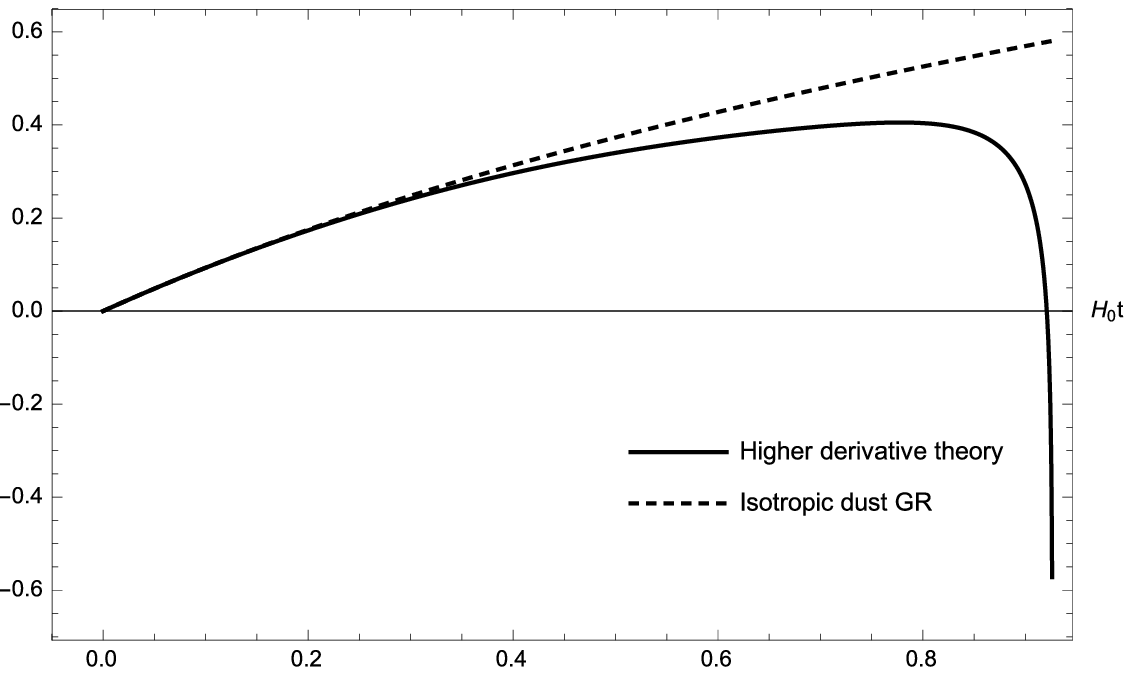}}
\subfigure[fig1b][anisotropies]
{\includegraphics[width=0.5\textwidth]{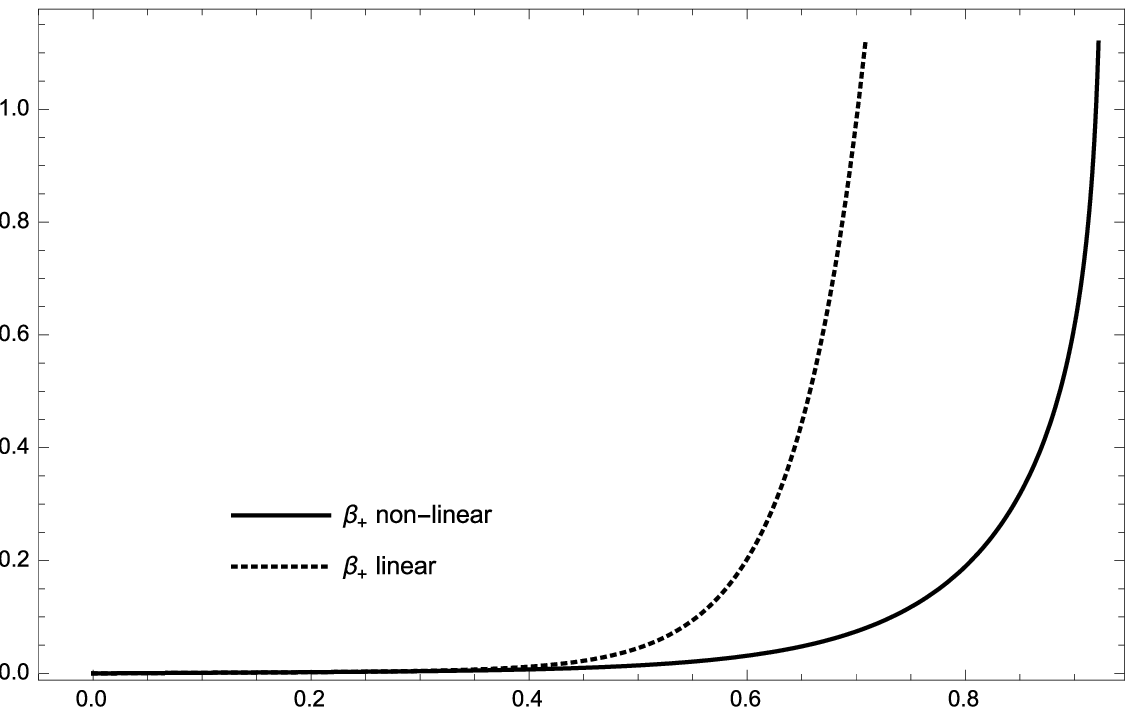}}
\begin{quotation}
\caption{For $\,a_{1}=+1\,$ and $\,a_{2}=1\,$ in the
matter-dominated case. The  tachyonic ghost instability is
qualitatively the same as in  the cosmological constant case,
confirming the correspondence with zero-frequencies limit of
the gravitational wave.}
\label{a1positive-dust}
\end{quotation}
\end{figure}
The examples of the results of numerical analysis can be seen in the
figures presented below. The qualitative behavior is pretty much the
same for any choice of initial data which we tried. The values for
the plots which we selected are specified at the Captions of the
figures. In all cases, the initial conditions for $\,\be_{+}(\tau)\,$ for
both linear and non-linear equations which we show in the plots are
$\,\be_\pm(0)\,=\,0$, \ $\dot{\be_\pm}(0)\,=\,0.01$,
\ $\ddot{\be_\pm}(0)\,=\,-0.001$, \ $\be_\pm^{(3)}(0)\,=\,0.0001$.
Furthermore in order to shorten the numerical procedure,
the value of \HL parameter has been taken as
$\,H_{0}\,=\,10^{-2}\,M_{p}$.
In the figures we present the plots of numerical solutions for
$\,\si(\tau)\,$ and anisotropies. In the last case we show only
$\,\be_{+}(\tau)\,$ solutions, because it turns out that both
anisotropies $\,\be_\pm(\tau)\,$ have similar behaviour, which
may differ only due to the choice of initial conditions and do not
define the asymptotic behaviour. The time $\tau$ is measured
in units of $1/H_0$, where we choose $H_0=0.01 M_P$ for the
sake of convenience of numerical analysis and plotting the
figures.

\begin{figure}[H]
\centering
 \subfigure[fig1a][$\sigma$ solutions]
{\includegraphics[width=0.5\textwidth]{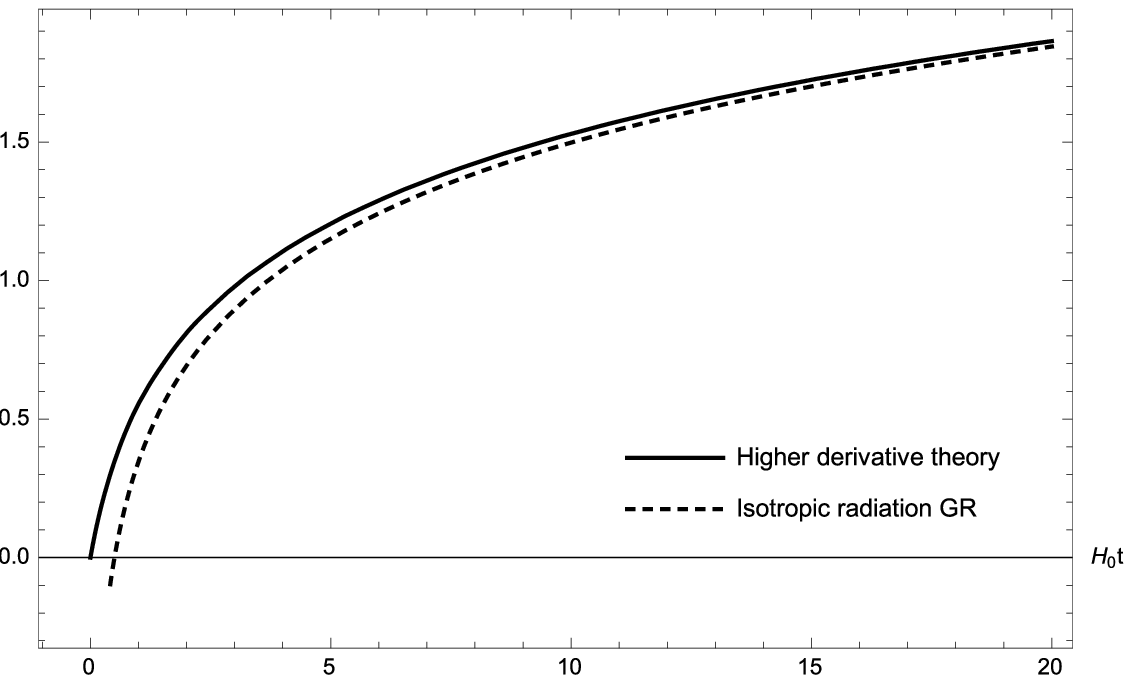}}
\qquad
\subfigure[fig1b][anisotropies]
{\includegraphics[width=0.5\textwidth]{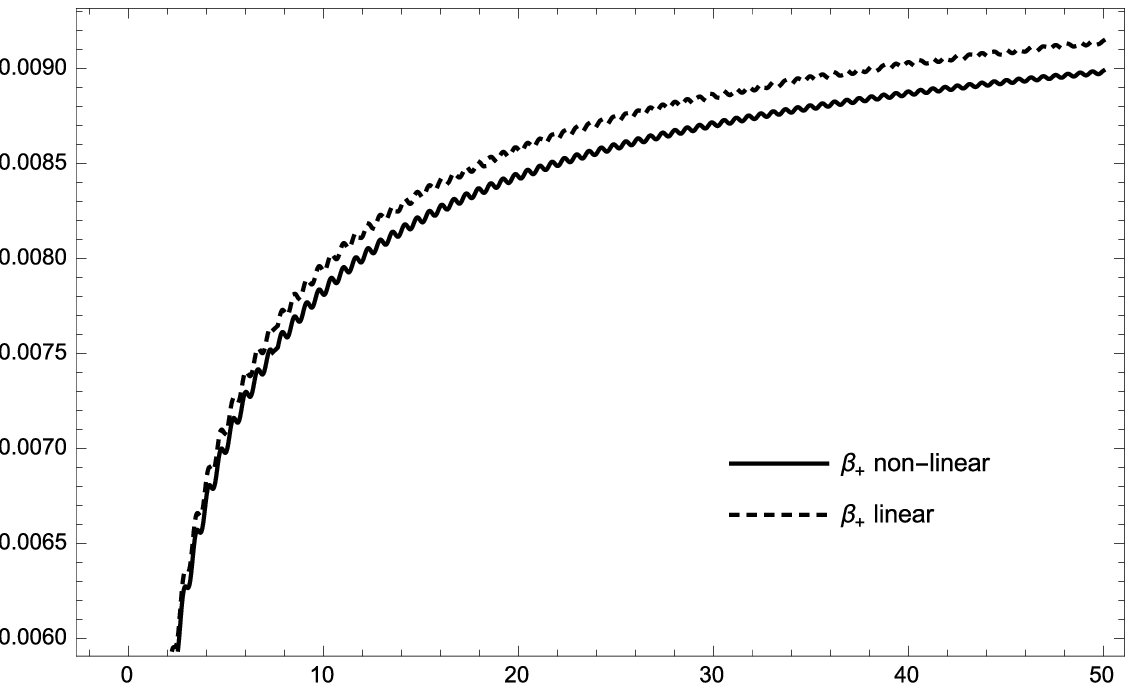}}
\begin{quotation}
\caption{For $\,a_{1}=-1\,$ and $\,a_{2}=1\,$ case we compare
the plots of $\,\si(\tau)\,$ and anisotropies from numerical solution
on the background  of  isotropic radiation - dominated solution of
general relativity.}
\label{radiationvalores1}
\end{quotation}
\end{figure}

\begin{figure}[!h]
\centering
 \subfigure[fig1a][$\sigma$ solutions]
{\includegraphics[width=9cm, height = 3.9cm]{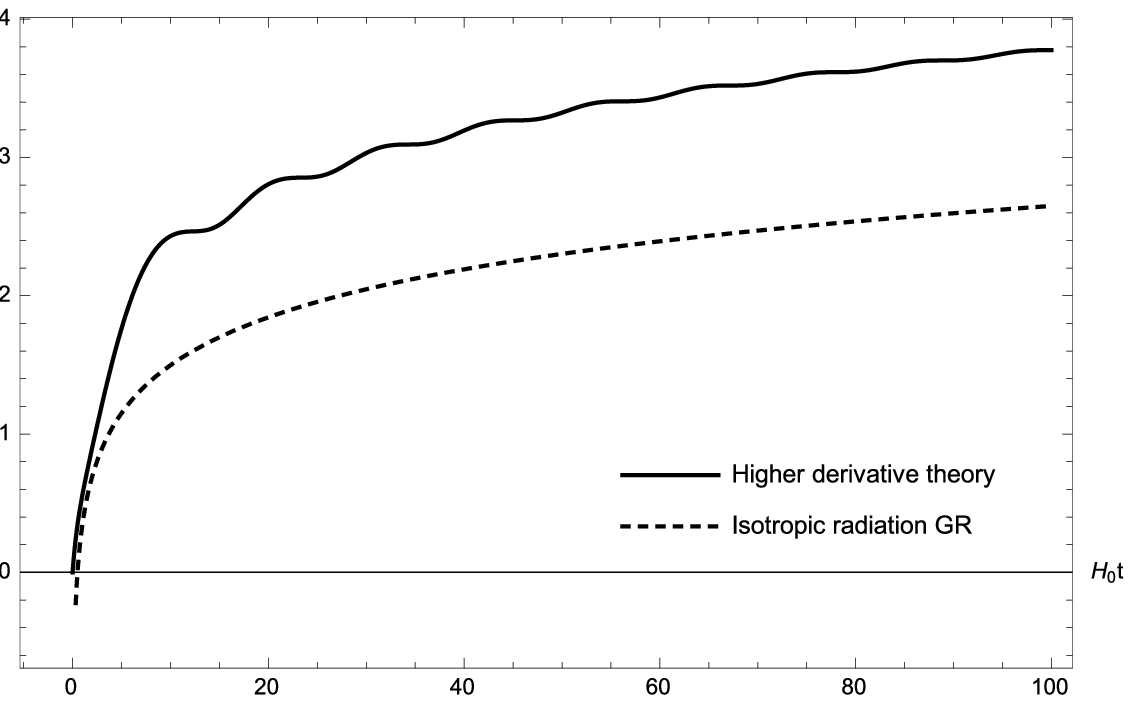}}
\subfigure[fig1b][anisotropies]
{\includegraphics[width=8.5cm, height = 3.9cm]{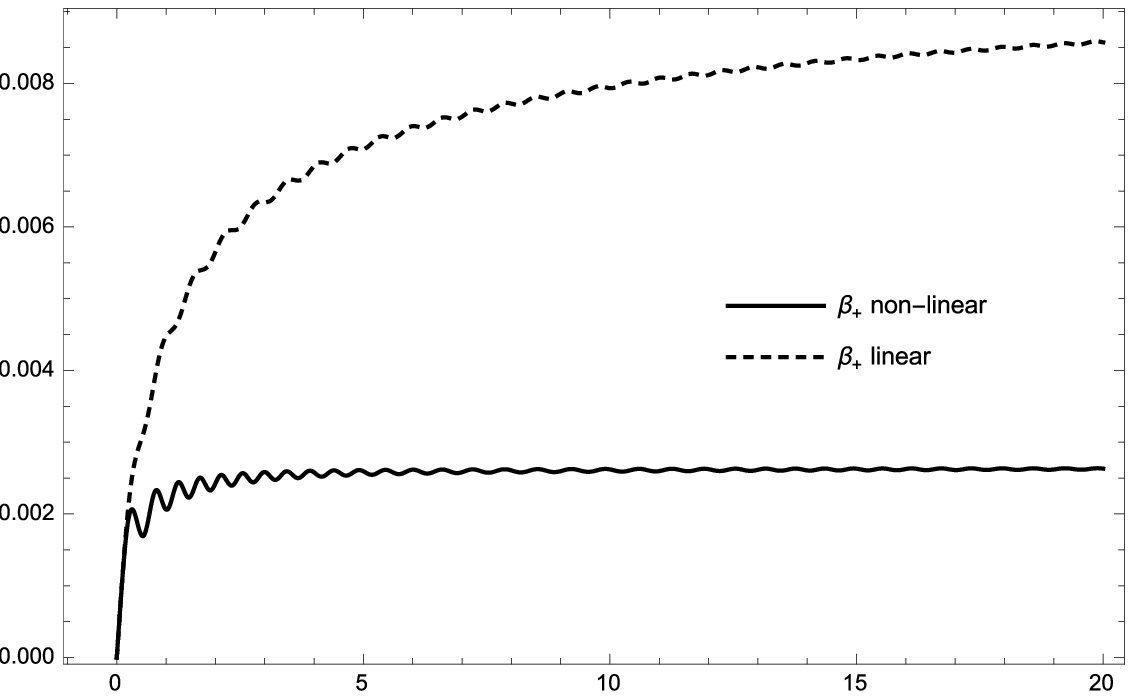}}
\begin{quotation}
\caption{The same plots as in Fig.~\ref{radiationvalores1}, but for the
different parameters $\,a_{1}=-1\,$ and $\,a_{2}=100$. This shows the
changes due to the large $R^2$-term, which is typical for the
Starobinsky inflation \cite{star,star83}. }
\label{radiationvalores2}
\end{quotation}
\end{figure}

\begin{figure}[!h]
\centering
 \subfigure[fig1a][$\sigma$ solutions]
{\includegraphics[width=9cm, height = 3.9cm]{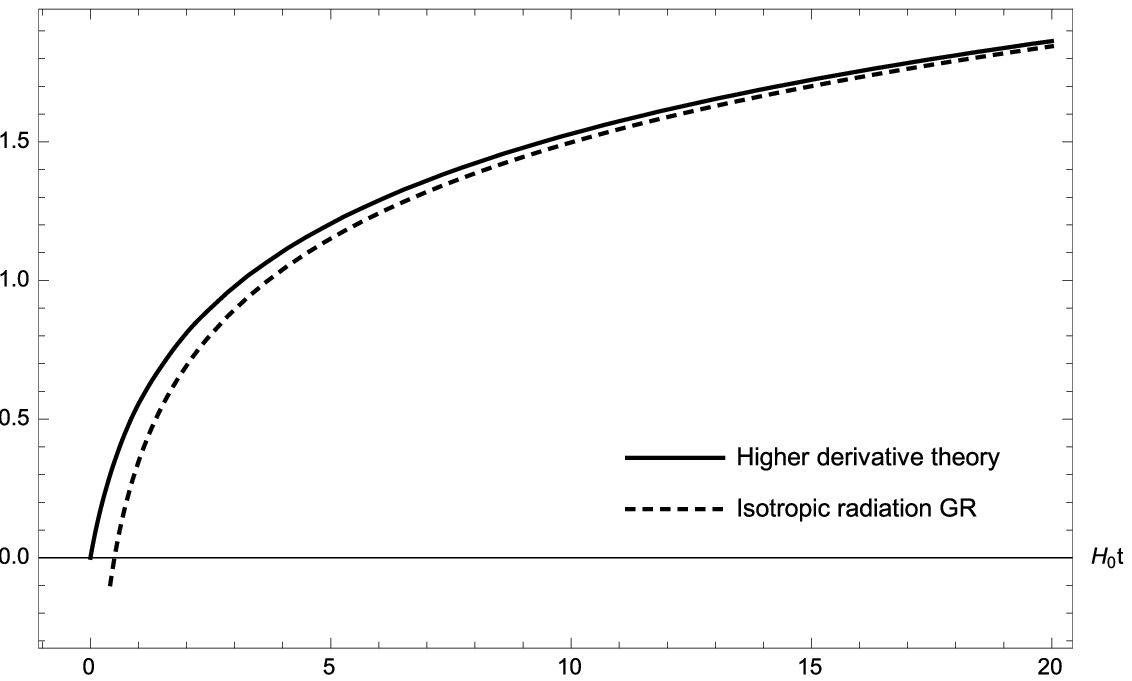}}
\subfigure[fig1b][anisotropies]
{\includegraphics[width=8.5cm, height = 3.9cm]{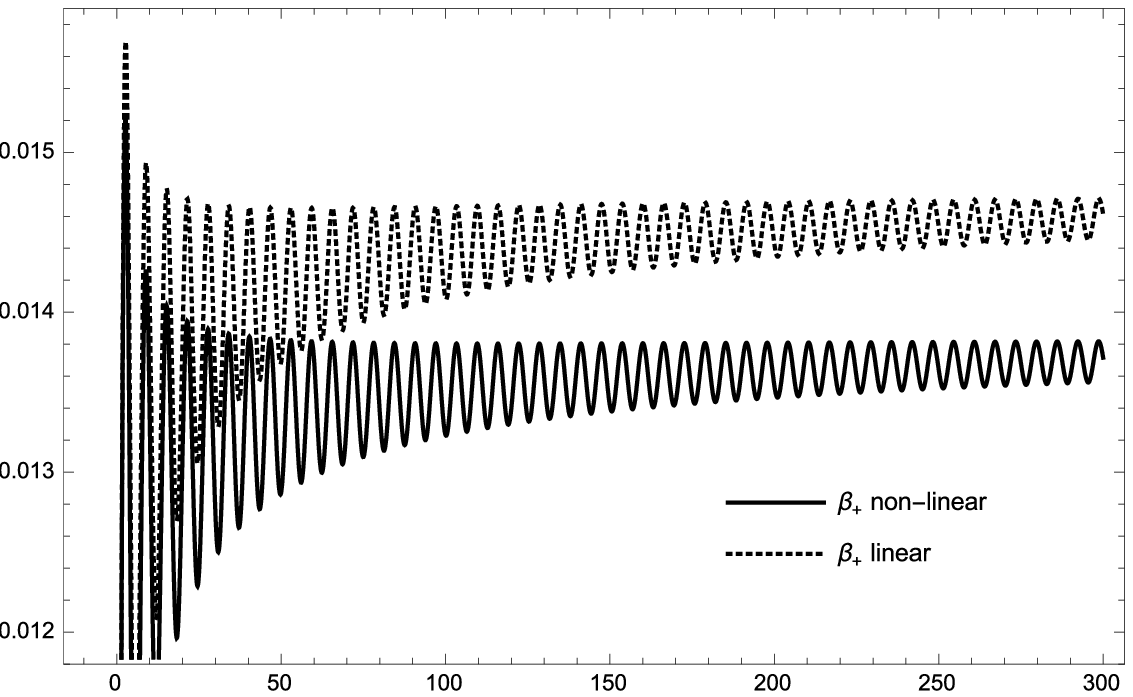}}
\begin{quotation}
\caption{The same plots, but with the large Weyl-squared term,
$\,a_{1}=-100\,$ and $\,a_{2}=1$.}
\label{radiationvalores3}
\end{quotation}
\end{figure}

In the first set, illustrated in Figs.~\ref{radiationvalores1},
\ref{radiationvalores2} and \ref{radiationvalores3}
the system of nonlinear equations have initial
conditions for $\,\si(\tau)\,$ which are the same as for isotropic
radiation - dominated universe in general relativity. Linearization is
done around $\,\si_{0}(\tau)\,$ of isotropic radiation dominated
universe.

\begin{figure}[!h]
\centering
 \subfigure[fig1a][$\sigma$ solutions]
{\includegraphics[width=9cm, height = 3.9cm]{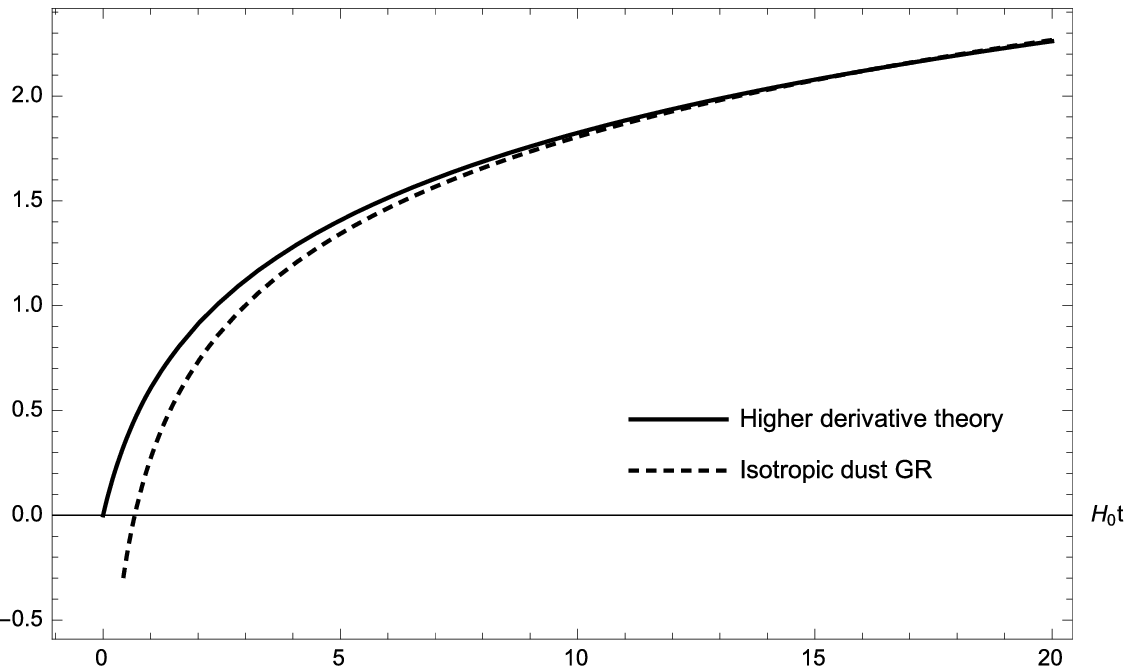}}
\qquad
\subfigure[fig1b][anisotropies]
{\includegraphics[width=8.5cm, height = 3.9cm]{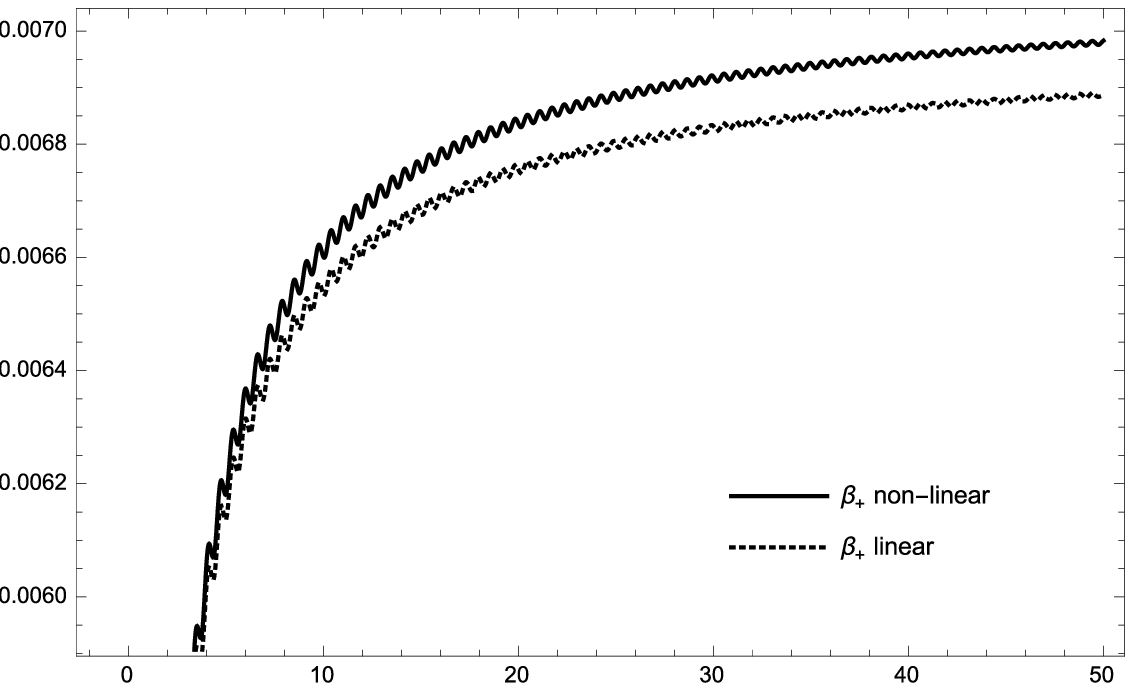}}
\begin{quotation}
\caption{The plots for the values $\,a_{1}=-1\,$ and $\,a_{2}=1\,$
with the background of isotropic matter - dominated solutions of
general relativity.}
\label{dustvalores1}
\end{quotation}
\end{figure}

\begin{figure}[H]
\centering
 \subfigure[fig1a][$\sigma$ solutions]
{\includegraphics[width=9cm, height = 3.9cm]{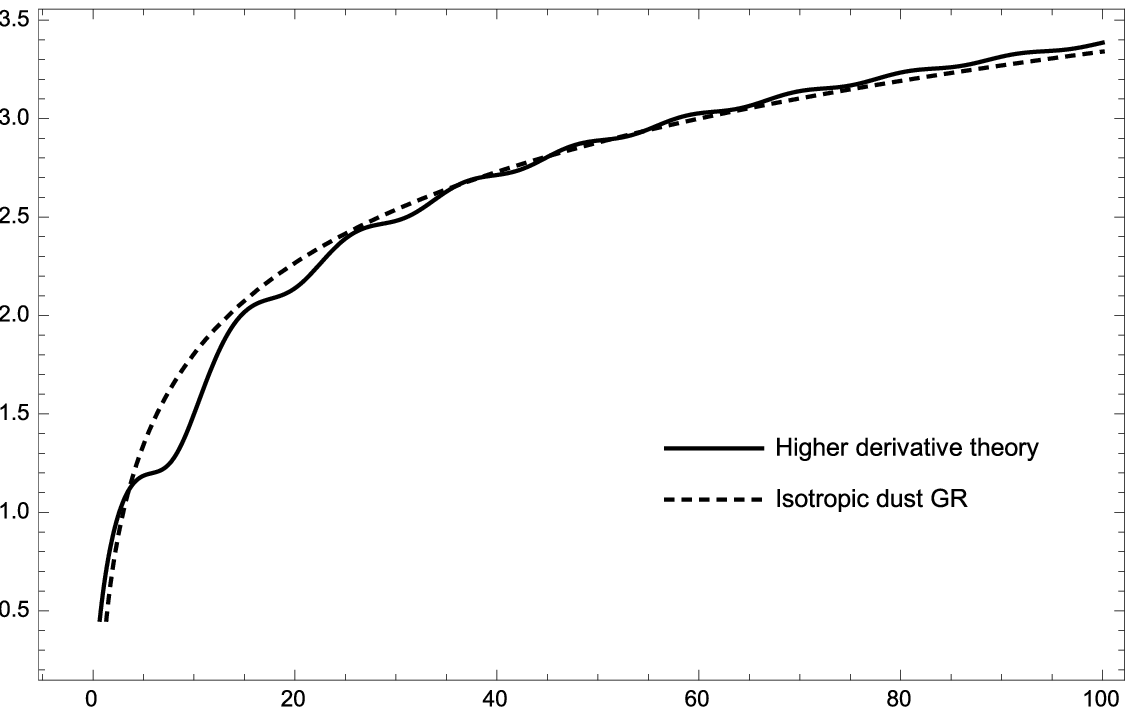}}
\qquad
\subfigure[fig1b][anisotropies]
{\includegraphics[width=8.5cm, height = 3.9cm]{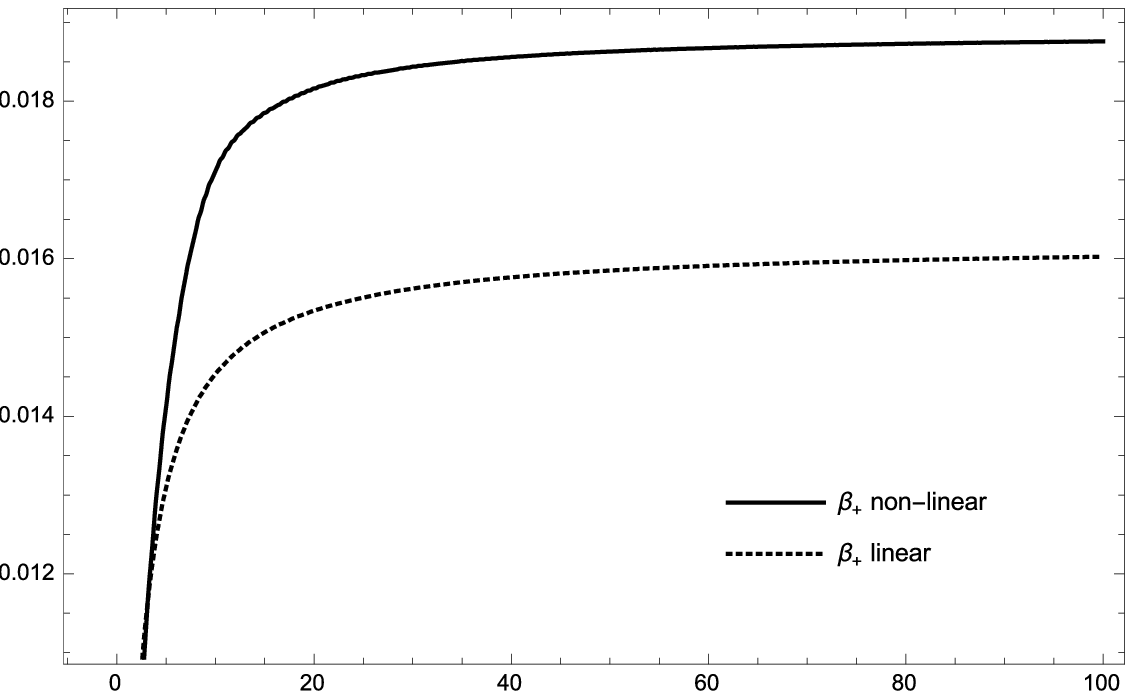}}
\begin{quotation}
\caption{The same as Fig.~\ref{dustvalores1}, but with the values
 $\,a_{1}=-1\,$ and $\,a_{2}=100\,$, intended to illustrate the effect
 of large $R^2$ term in the Starobinsky inflation. The background is
 dominated by dust.}
\label{dustvalores2}
\end{quotation}
\end{figure}

\begin{figure}[H]
\centering
 \subfigure[fig1a][$\sigma$ solutions]
{\includegraphics[width=9cm, height = 4.3cm]{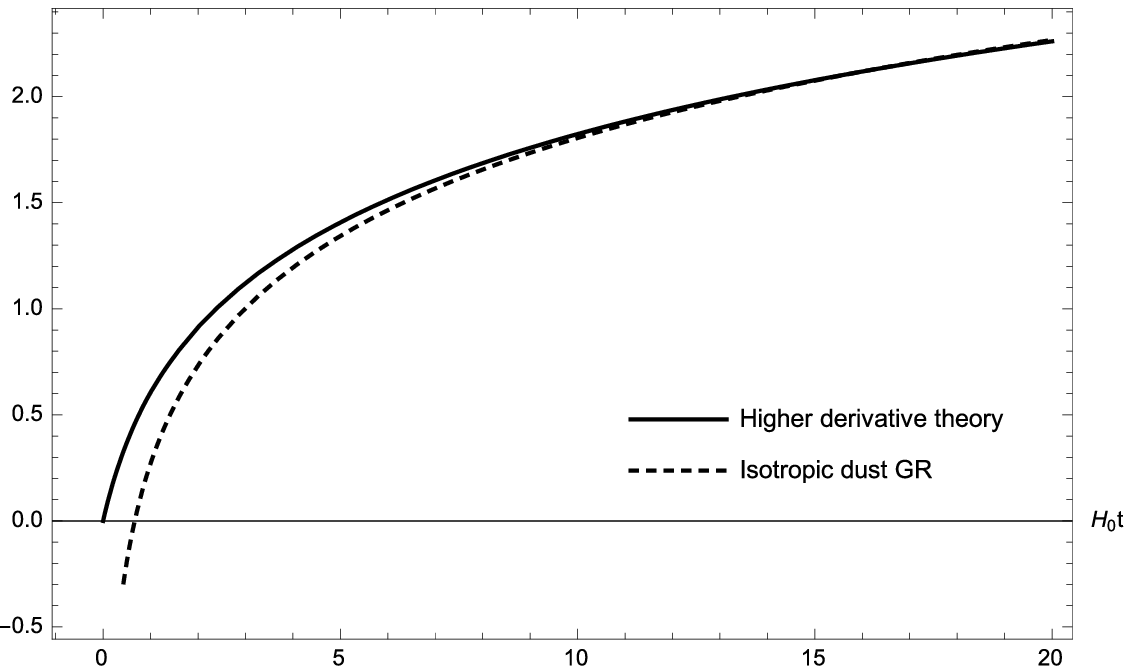}}
\qquad
\subfigure[fig1b][anisotropies]
{\includegraphics[width=8.5cm, height = 4.3cm]{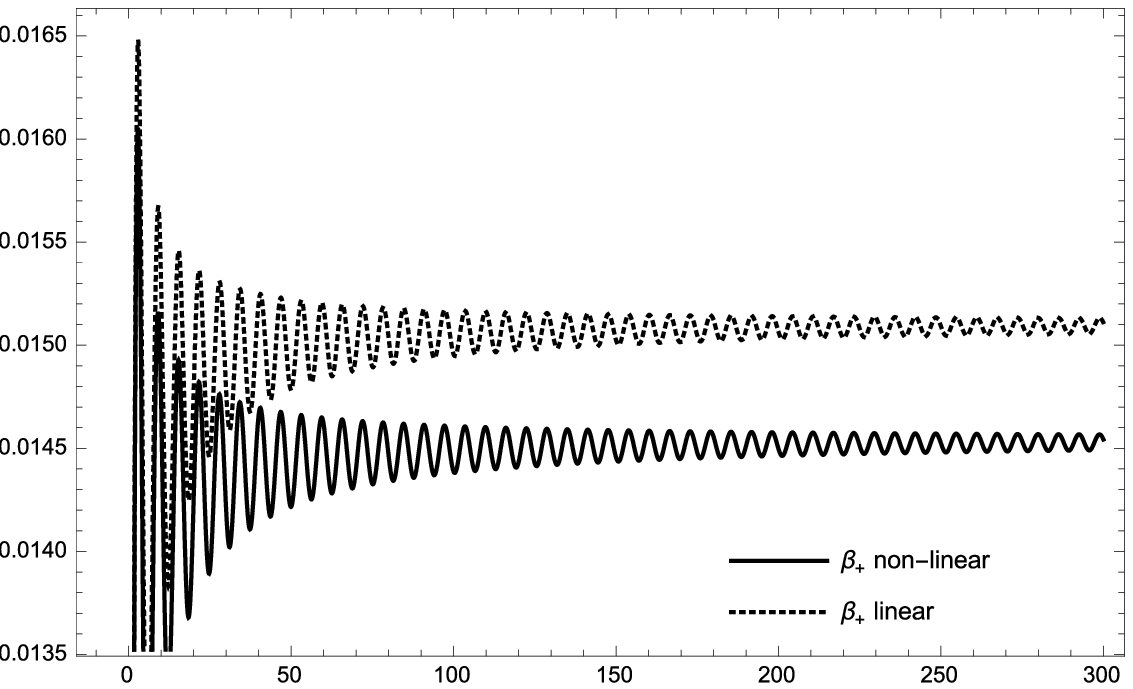}}
\begin{quotation}
\caption{The same of Fig.~\ref{dustvalores1}, but with the values
$\,a_{1}=-100\,$ and $\,a_{2}=1$.}
\label{dustvalores3}
\end{quotation}
\end{figure}

The second set of Figs.~\ref{dustvalores1}, \ \ref{dustvalores2}
and \ref{dustvalores3}  illustrates the solutions for the background
of $\,\si_{0}(\tau)\,$ corresponding to the matter - dominated
universe.

\begin{figure}[H]
\centering
 \subfigure[fig1a][$\sigma$ solutions]
{\includegraphics[width=8cm, height = 4.0cm]{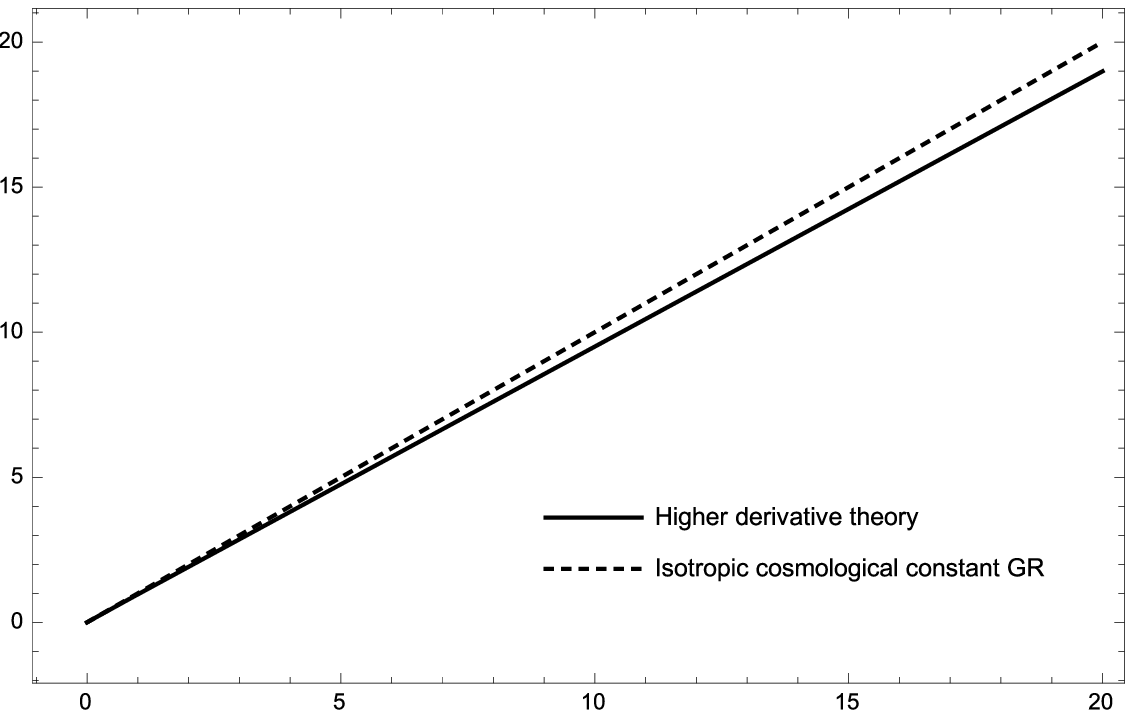}}
\qquad
\subfigure[fig1b][anisotropies]
{\includegraphics[width=8cm, height = 4.0cm]{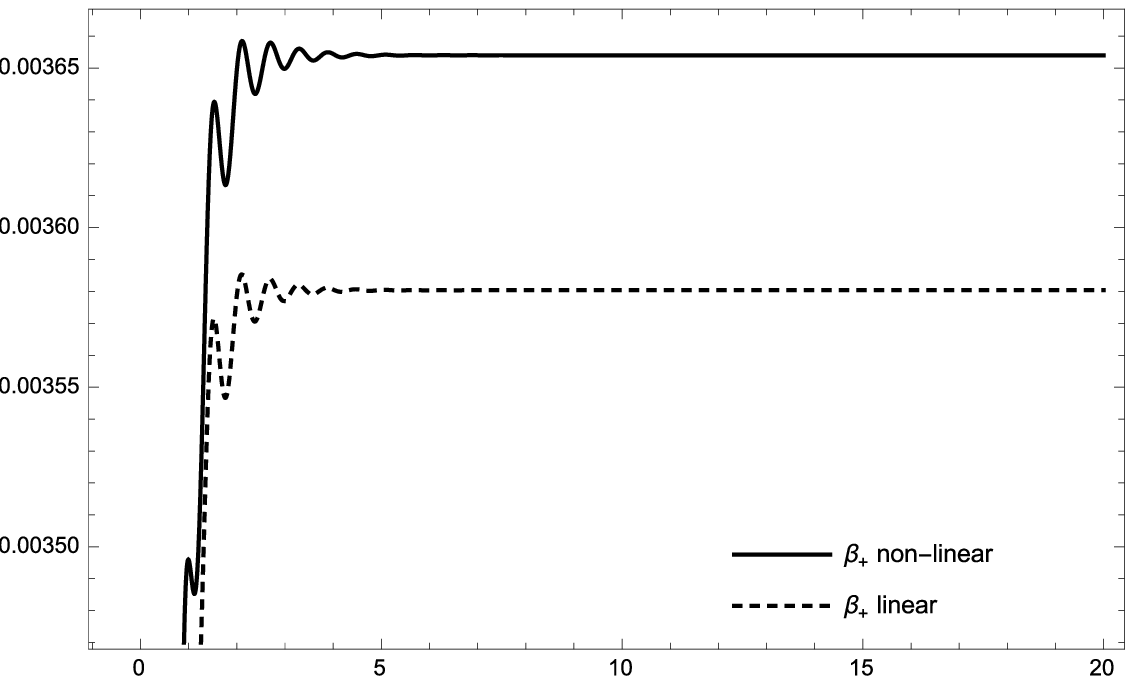}}
\begin{quotation}
\caption{The plots for $\,a_{1}=-1\,$ and $\,a_{2}=1\,$,
 for equations on the isotropic cosmological constant - dominated
 background.}
\label{ccvalores1}
\end{quotation}
\end{figure}

\begin{figure}[H]
\centering
 \subfigure[fig1a][$\sigma$ solutions]
{\includegraphics[width=9.0cm, height = 4.2cm]{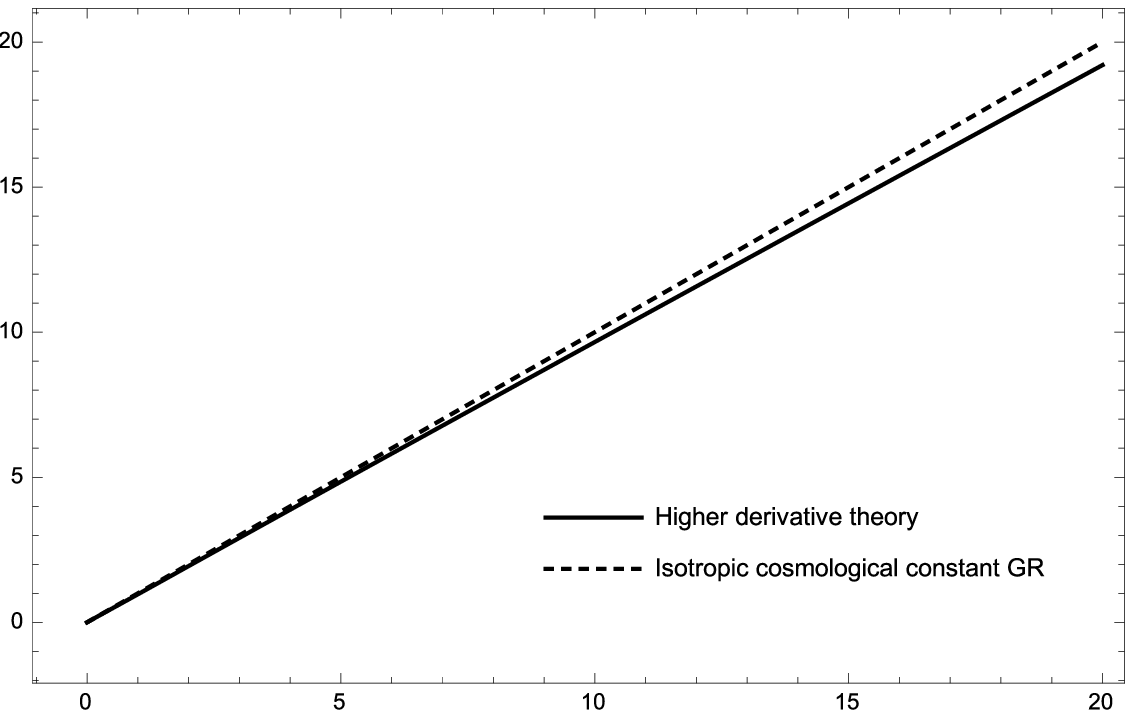}}
\qquad
\subfigure[fig1b][anisotropies]
{\includegraphics[width=9cm, height = 4.2cm]{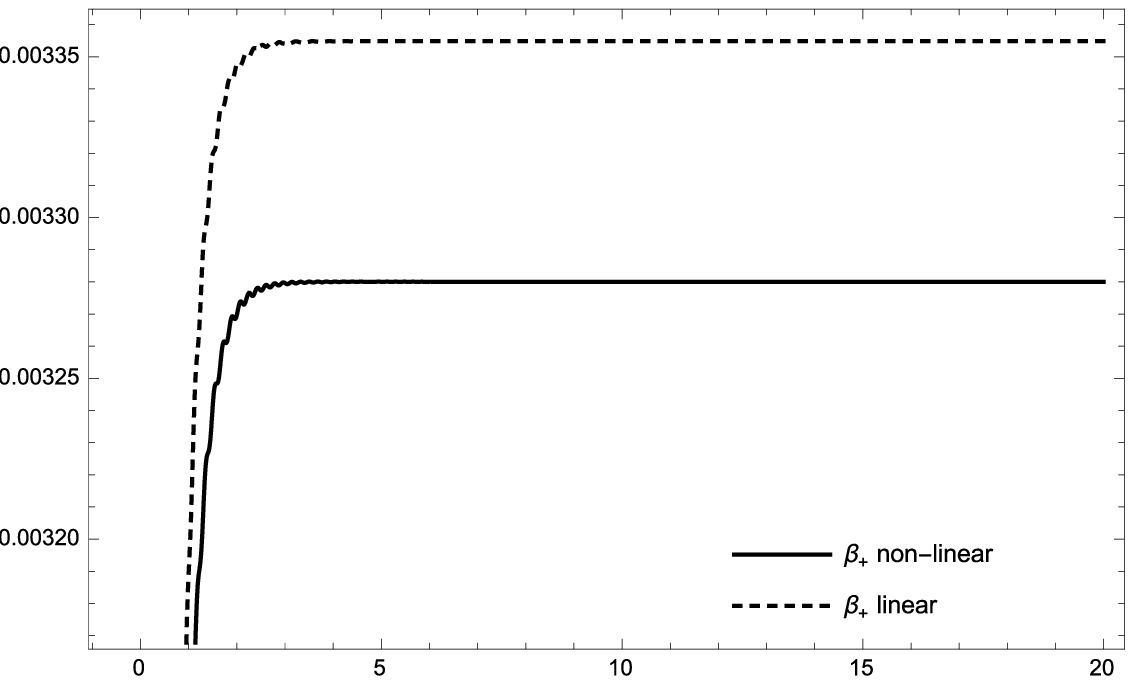}}
\begin{quotation}
\caption{The same as Fig.~\ref{ccvalores1}, but with the values
$\,a_{1}=-1\,$ and $\,a_{2}=100$.}
\label{ccvalores2}
\end{quotation}
\end{figure}

\begin{figure}[H]
\centering
 \subfigure[fig1a][$\sigma$ solutions]
{\includegraphics[width=9cm, height = 3.9cm]{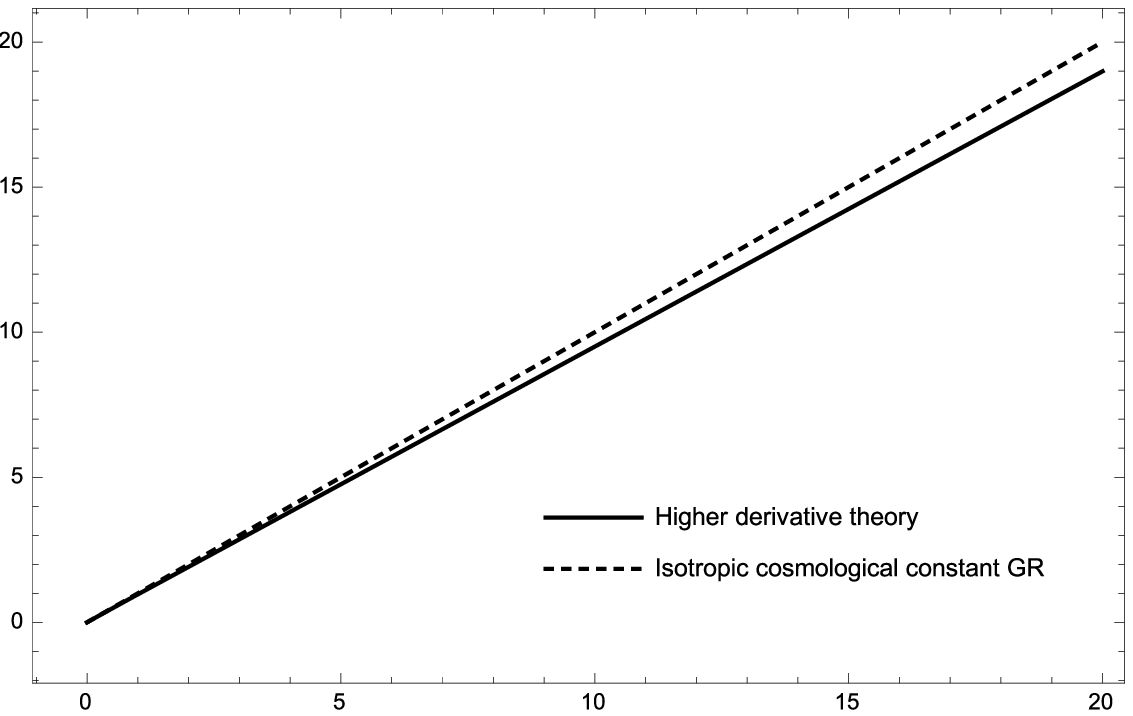}}
\qquad
\subfigure[fig1b][anisotropies]
{\includegraphics[width=9cm, height = 4.4cm]{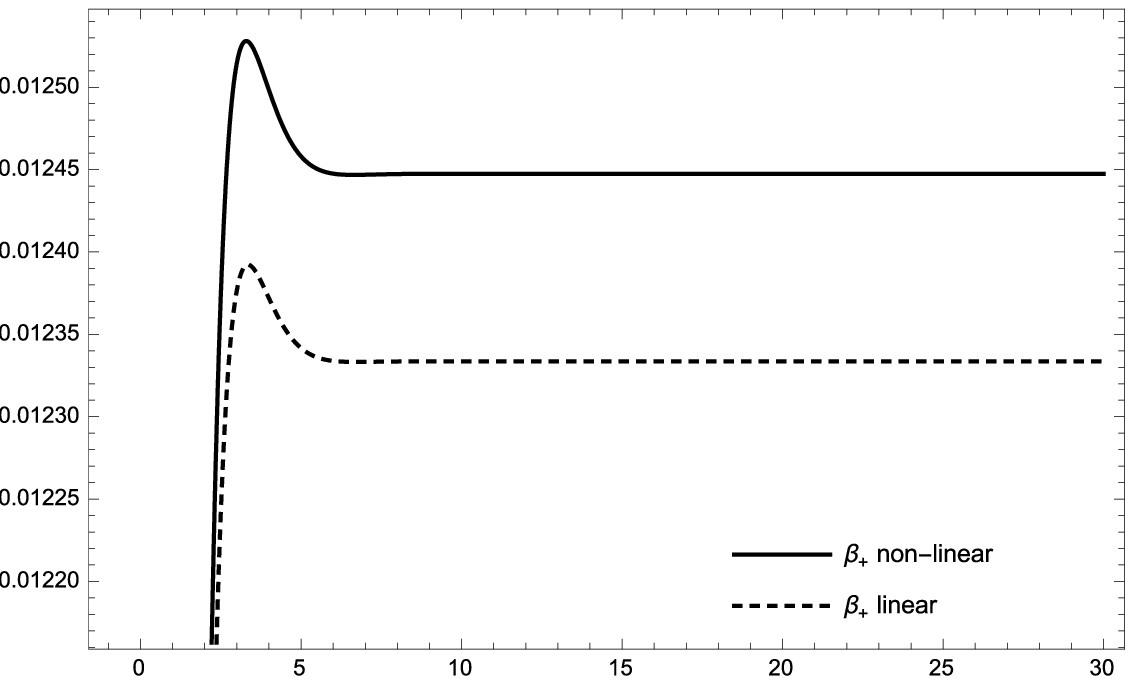}}
\begin{quotation}
\caption{The same as Fig.~\ref{ccvalores1}, but with the values
$\,a_{1}=-100\,$ and $\,a_{2}=1,$.}
\label{ccvalores3}
\end{quotation}
\end{figure}

The last cases are shown in Figs.~\ref{ccvalores1},
\ref{ccvalores2} and \ref{ccvalores3}, they correspond to
equations for the variation of conformal factor and anisotropies
on the background of isotropic solution in the universe dominated
by cosmological constant.
Let us conclude this section by repeating that we have also checked
other choices of initial data and the results are always qualitatively
the same as in the plots shown above. In general there is a very good
correspondence between linearized \B1 system and the dynamics
of gravitational waves with low frequencies from one side, and the
linearized and non-perturbative treatments from another side.
\section{Conclusions}
\label{S4}

We have explored the time dependence of anisotropies in the \B1 model
with fourth derivatives, which can be seen as a zero-frequency
approximation for the gravitational waves in the model (\ref{action}).
Qualitatively we observe from the plots presented in the Figures
that in all cases there is no qualitative difference between the behaviour
of linearized and non-perturbative systems, exactly as it should be in
accordance with the standard mathematical results cited in
Sec.~\ref{Smath}.

In all cases which we were analysed, the dynamics of both linearized
and general systems does not show instabilities related to the presence
of higher derivatives, exactly as one should expect from the previous
considerations of the gravitational waves from one side
\cite{HD-Stab} and the
mentioned mathematical theorems from another side. Since \B1  can
be regarded as a zero-frequency approximation to the gravitational
waves dynamics, we gain a strong reasons to expect the absence of
explosive exponential type instabilities for the gravitational
waves, even in the nonperturbative regime.

For the cases of radiation-dominated and dust-dominated background
solutions the numerical results confirm show that for the values
$\,a_1\,=\,-1\,$ and $\,a_2\,=\,1\,$ the numerical solutions of
$\,\si(\tau)\,$ asymptotically tend to the isotropic ones with the same
matter contents. At the same time, for larger value $\,a_2\,=\,100\,$
we can note stronger deviation between linear and nonperturbative
regimes. This effect should be expected much stronger for the
phenomenologically optimized value $\,a_2\,\approx \,5 \times 10^8$,
required for the successful Starobinsky inflation \cite{star,star83}.

In general, we confirmed the expectations of \cite{ HD-Stab} and
\cite{PP} concerning the correspondence between linear and general
nonlinear results. It would be certainly interesting to extend the
analysis in several directions. For instance, to include the cases of
the background cosmological metrics with strong curvature, such
that the effect of higher derivatives on the background should be
taken into account. Regardless of that this case is not expected to
give great surprises (the reason is that the large $a_2$ is known to
increase the value of $H_0$, in the first approximation), this check
has to be done. In fact, the solutions for more complicated
cosmological backgrounds would be an interesting issue to explore.
A much more challenging problem is to consider more complicated
anisotropic solutions, with a non-zero frequencies. Such an
investigation would require more serious calculation, but in some
cases it does not look impossible. Anyway, the results of the
present work show that we have strong reasons to believe to the
validity of the first-order perturbations if they show the strong
signs of asymptotic stability.
\section*{Acknowledgements}
S.C.R. is grateful to CAPES for supporting his Ph.D. project.
I.Sh. was partially supported by
CNPq (grant 303893/2014-1) and FAPEMIG (project APQ-01205-16).


\end{document}